\documentclass{article}
\usepackage{FPSdev}
\usepackage{epsfig}
\newcommand{\vev}{\ensuremath{<\phi>_0}}
\newcommand{\fbi}{\ensuremath{\mathrm{fb}^{-1}}}
\newcommand{\pbi}{\ensuremath{\mathrm{pb}^{-1}}}
\newcommand{\LEPI}  {\mbox{LEP1}}
\newcommand{\LEPII} {\mbox{LEP2}}
\newcommand{\RunI}  {\mbox{Run I}}
\newcommand{\RunII} {\mbox{Run II}}
\newcommand{\RunIIa}{\mbox{Run IIa}}

\newcommand{\WW}{\ensuremath{\mathrm{W}^+\mathrm{W}^-}}
\newcommand{\epem}{\ensuremath{\mathrm{e}^+\mathrm{e}^-}}
\newcommand{\eetoWW}{\ensuremath{\epem\rightarrow\WW}}
\newcommand{\qqWln}{\ensuremath{\qq\rightarrow\mathrm{W}\rightarrow\lnu}}
\newcommand{\roots}{\ensuremath{\sqrt{s}}}

\newcommand{\Mw}{\ensuremath{M_{\mathrm{W}}}}
\newcommand{\Gw}{\ensuremath{\Gamma_{\mathrm{W}}}}
\newcommand{\Mz}{\ensuremath{M_{\mathrm{Z}}}}
\newcommand{\Gz}{\ensuremath{\Gamma_{\mathrm{Z}}}}
\newcommand{\Mh}{\ensuremath{M_{\mathrm{H}}}}
\newcommand{\Mt}{\ensuremath{M_{\mathrm{t}}}}
\newcommand{\sigW}{\ensuremath{\sigma_{\mathrm{W}}}}
\newcommand{\sigZ}{\ensuremath{\sigma_{\mathrm{Z}}}}
\newcommand{\sinw}{\ensuremath{\sin^2\!\theta_{\mathrm{W}}}}
\newcommand{\tanw}{\ensuremath{\tan^2\!\theta_{\mathrm{W}}}}
\newcommand{\CC}{\mbox{{\sc CC03}}}

\newcommand{\mt}{\ensuremath{m_{\mathrm{T}}}}
\newcommand{\pt}{\ensuremath{p_{\mathrm{T}}}}
\newcommand{\ptw} {\ensuremath{\pt^{\, \mathrm{W}}}}

\newcommand{\pte} {\ensuremath{\pt^{\, e}}}
\newcommand{\ptnu}{\ensuremath{\pt^{\, \nu}}}
\newcommand{\Et}{\ensuremath{E_{\mathrm{T}}}}

\newcommand{\Etl} {\ensuremath{\Et^{\, \ell}}}
\newcommand{\Etnu}{\ensuremath{\Et^{\, \nu}}}
\newcommand{\vecptl} {\ensuremath{\vec{p}_{\mathrm{T}}^{\, \ell}}}
\newcommand{\vecptnu} {\ensuremath{\vec{p}_{\mathrm{T}}^{\, \nu}}}
\newcommand{\vecU} {\ensuremath{\vec{U}}}
\newcommand{\ppbar}{\ensuremath{\mathrm{p}\bar{\mathrm{p}}}}
\newcommand{\stat}{\mathrm{(stat.)}}
\newcommand{\syst}{\mathrm{(syst.)}}
\newcommand{\Br}{\ensuremath{\mathrm{Br}}}
\newcommand{\Wtomn}{\mbox{$\mathrm{W}\rightarrow\mnu$}}
\newcommand{\Wtoen}{\mbox{$\mathrm{W}\rightarrow\enu$}}
\newcommand{\Wtotn}{\mbox{$\mathrm{W}\rightarrow\tnu$}}
\newcommand{\Wtoln}{\mbox{$\mathrm{W}\rightarrow\lnu$}}
\newcommand{\Wtoqq}{\mbox{$\mathrm{W}\rightarrow\qq$}}
\newcommand{\Ztoee}{\mbox{$\mathrm{Z}\rightarrow\ee$}}
\newcommand{\Zz}       {\ensuremath{{\mathrm{Z}^0}}}

\newcommand{\Zll}      {\ensuremath{\Zz\rightarrow\lplm}}
\newcommand{\Zqq}      {\ensuremath{\Zz\rightarrow\qq}}
\newcommand{\nue}   {\ensuremath{\nu_{e}}}
\newcommand{\numu}  {\ensuremath{\nu_{\mu}}}
\newcommand{\nutau} {\ensuremath{\nu_{\tau}}}
\newcommand{\nuell} {\ensuremath{\nu_{\ell}}}
\newcommand{\enu}   {\ensuremath{\mathrm{e}\nue}}
\newcommand{\mnu}   {\ensuremath{\mu\numu}}
\newcommand{\tnu}   {\ensuremath{\tau\nutau}}
\newcommand{\lnu}   {\ensuremath{\ell{\nuell}}}

\newcommand{\qq}    {\ensuremath{\mathrm{q\overline{q}}}}
\newcommand{\ee}    {\ensuremath{ee}}
\newcommand{\lplm}  {\ensuremath{\ell^+\ell^-}}

\newcommand{\qqen}{\ensuremath{\qq\enu}}

\newcommand{\qqtn}{\ensuremath{\qq\tnu}}

\newcommand{\lnlns}{\ensuremath{\ell\nu\ell\nu}}
\newcommand{\qqlns}{\ensuremath{\mathrm{qq}\ell\nu}}
\newcommand{\qqqqs}{\ensuremath{\mathrm{qqqq}}}
\newcommand{\Vij} {\mbox{$|\mathrm{V}_{ij}|$}}
\newcommand{\Vud} {\mbox{$|\mathrm{V}_{\mathrm{ud}}|$}}
\newcommand{\Vus} {\mbox{$|\mathrm{V}_{\mathrm{us}}|$}}
\newcommand{\Vcd} {\mbox{$|\mathrm{V}_{\mathrm{cd}}|$}}
\newcommand{\Vcb} {\mbox{$|\mathrm{V}_{\mathrm{cb}}|$}}
\newcommand{\Vub} {\mbox{$|\mathrm{V}_{\mathrm{ub}}|$}}
\newcommand{\Vcs} {\mbox{$|\mathrm{V}_{\mathrm{cs}}|$}}
\newcommand{\Rcw} {\ensuremath{R_{c}^W}}
\newcommand{\WWg} {\ensuremath{\mathrm{WW}\gamma}}
\newcommand{\WWgg} {\ensuremath{\mathrm{WW}\gamma\gamma}}
\newcommand{\WWZg} {\ensuremath{\mathrm{WWZ}\gamma}}
\newcommand{\kg}{\ensuremath{\kappa_\gamma}}
\newcommand{\kz}{\ensuremath{\kappa_\mathrm{Z}}}
\newcommand{\lag}{\ensuremath{\lambda_\gamma}}
\newcommand{\laz}{\ensuremath{\lambda_\mathrm{Z}}}
\newcommand{\gz}{\ensuremath{g_1^\mathrm{Z}}}
\newcommand{\dkg}{\ensuremath{\Delta\kg}}
\newcommand{\dkz}{\ensuremath{\Delta\kz}}
\newcommand{\dgz}{\ensuremath{\Delta\gz}}
\newcommand{\JETSET}{\mbox{J{\sc etset}}}

\newcommand{\ARIADNE}{\mbox{A{\sc riadne}}}
\newcommand{\GENTLE}{\mbox{G{\sc entle}}}

\newcommand{\HERWIG}{\mbox{H{\sc erwig}}}

\newcommand{\lnplusg} {\ensuremath{\lnu+\gamma}}
\newcommand{\pieeee}  {\ensuremath{\pi^0\rightarrow eeee}}
\newcommand{\Jee}      {\ensuremath{\mathrm{J}/\Psi\rightarrow ee}}
\newcommand{\Zee}      {\ensuremath{\Zz\rightarrow ee}}
\newcommand{\etal}    {\mbox{{\it et al.}}}
\newcommand{\Journal}[4] {{#1} \textbf{#2} {#4} (#3)}
\newcommand{\PL}  {Phys. Lett. }
\newcommand{\ZP}  {Z. Phys.}
\newcommand{\EPJ} {Eur. Phys. J.}

\newcommand{\PRL} {Phys. Rev. Lett.}
\newcommand{\PR}  {Phys. Rev.}

\newcommand{\CPC} {Comp. Phys. Comm.}

\begin{document}
\title{W BOSON PROPERTIES}
\author{
  Eric Torrence \\
 {\em Physics Department}\\
 {\em 1274 University of Oregon}\\
 {\em Eugene, Oregon 97403-1274, USA}\\
  \\
 {\em To appear in proceedings of $21^{\mathrm st}$ Physics in Collision}\\ 
 {\em Seoul, South Korea, 28--30 June 2001.}
}
\maketitle
\baselineskip=11.6pt
\begin{abstract}
Studying the properties of the W boson naturally plays
a key role in precision tests of the Standard Model.
In this paper, the key measurements performed at LEP
and the Tevatron over the last decade are reviewed.
The current world knowledge of the W boson production and 
decay properties, gauge couplings, and mass are presented,
with an emphasis on the most recent results from \LEPII.
Some estimates of the sensitivity of the upcoming
Tevatron \RunII\ are also presented.
\end{abstract}
\baselineskip=14pt
%
%
\section{Introduction}
At tree level, the electroweak sector of the Standard Model is completely
determined by three parameters, which are the gauge couplings to the
$SU(2)$ and $U(1)$ fields, and the vacuum expectation value of the Higgs field.
Experimentally, then, it only requires three precision measurements to
completely describe the theory, and all other parameters of the theory
can then be calculated.
The three most precise measurements available are the
electromagnetic coupling
$\alpha = \frac{1}{4\pi} \frac{(g_1 g_2)^2}{g_1^2+g_2^2}$ which is very well
known from electron $(g-2)$ experiments,
the Fermi coupling $G_F = \frac{1}{\sqrt{8}} \frac{1}{\vev}$ which
is measured with the Muon lifetime, and the mass of the Z boson
$\Mz = \frac{\vev}{\sqrt{2}} \sqrt{g_1^2 + g_2^2}$ measured at LEP.
Measurements of other electroweak parameters, most notably the 
mass of the W boson and the weak mixing angle, can
then be used to directly test the predictions of the theory.

Due to the presence of higher-order radiative corrections,
the simple tree-level predictions of the Standard Model
are modified, predominantly through loop corrections,
such that additional parameters are needed to fully
describe the theory.
The mass of the W boson, for example is predicted to be
\begin{math}
M^2_{\mathrm{W}} = \frac{\pi\alpha}{\sqrt{2} \sinw G_F}(1+\Delta r), 
\end{math}
where $\Delta r = f(m^2_t, \ln(\Mh))$.
These radiative corrections can either be seen as an annoying
complication of the model or as an opportunity to gain indirect
knowledge of otherwise unknown parameters like the Higgs mass.

Studying the properties of the W boson has been facilitated
in the last decade by the direct production of large numbers
of W bosons at LEP and the Tevatron. 
The LEP collider at CERN is an \epem\ collider which was operated
at the Z pole from 1990 -- 1995 (\LEPI), and above the W boson
pair-production threshold during the period 1996 -- 2000 (\LEPII).
The process \eetoWW\ can be identified with high efficiency
in all decay modes of the W boson, and with a cross section of 
around 17~pb at 200 GeV, the four LEP collaborations 
(Aleph, Delphi, L3, and Opal) collected
a total sample of around 35,000 W pairs in the 2.7~fb of data
recorded.

The Tevatron collider at Fermilab is a \ppbar\ collider which ran
at 1.8 TeV during 1992 -- 1995 (\RunI), delivering colliding
beams to the D0 and CDF collaborations.
In hadronic collisions, W bosons are identified in the leptonic
decay modes, with high \pt\ lepton triggers providing samples
of \Wtoen\ and \Wtomn\ decays.
With a very large production rate of 
$\sigma_{WX} \cdot\Br(\Wtoln) \approx 2.4$~nb,
the \RunI\ total W sample exceeded 200,000 events.
Starting in 2001, an upgraded Tevatron will begin \RunII,
which is expected to deliver 2 \fbi\ at 2~TeV by 2003 (\RunIIa),
and possibly up to 30 \fbi\ by the end of the decade.

Since the Tevatron has not been taking data for over five years,
most of the recent results in W physics have come from the LEP
collaborations, and these will be highlighted in this paper.
Where appropriate, the sensitivity of the Tevatron experiments
after data taken during the \RunIIa\ period will also be noted.
Unless otherwise specified, all results are preliminary.
%
%
%
\section{W Boson Production and Decay}
\label{sec:prod}
\subsection{W-pair production at \LEPII}
\begin{figure}[htb]
  \begin{center}
    \epsfig{file=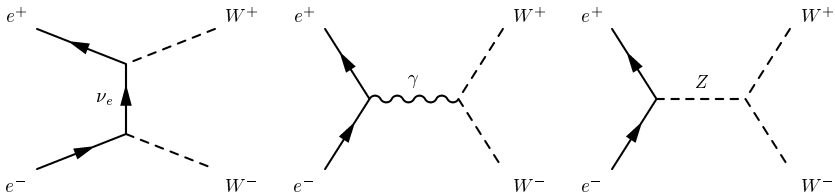,width=11cm}
    \caption{\it
      The \CC\ diagrams for W-pair production.
      \label{fig:CC03}
      }
  \end{center}
\end{figure}

In this paper, \WW\ events produced in \epem\ collisions are defined
in terms of the \CC\ class of production diagrams shown in 
Figure~\ref{fig:CC03} following the notation of the LEP 
collaborations\cite{bib:LEP2YR}.
These amplitudes
provide a natural definition of resonant
W-pair production, even though other non-\CC\ diagrams contribute to the
same four-fermion final states.
In order to correctly account for the additional non-\CC\ amplitudes
present in the production process, the LEP collaborations use Monte
Carlo generators based on complete four-fermion 
calculations\cite{bib:KORALW,bib:EXCALIBUR,bib:GRC4F}.
The difference in production rate between purely \CC\ diagrams and
the complete four-fermion matrix elements is typically less than
5\% depending upon the final state involved.

In the Standard Model, \WW\ events are expected to decay into fully
leptonic (\lnlns), semi-leptonic (\qqlns), 
or fully hadronic (\qqqqs) final states with predicted 
branching fractions of 10.6\%, 43.9\%, and 45.6\% 
respectively\cite{bib:LEP2YR}.
The LEP collaborations use separate selections to identify these
three main topologies, and the \lnlns\ and \qqlns\ events
are further classified according to the observed lepton type.
In total, \WW\ candidate events are selected in 
one of ten possible final states $(6\times\lnlns$, $3\times\qqlns$,
and $1\times\qqqqs)$.
Due to the clean environment of \epem\ collisions, all decay modes
of the \WW\ events can be reconstructed with good efficiency and
modest backgrounds as shown in Table~\ref{tab:wwsel}.

\begin{table}
  \centering
  \caption{\it \eetoWW\ selection summary.
    }
  \vskip 0.1 in
  \begin{tabular}{|l|c|c|c|} \hline
    Channel &  SM Rate & Efficiency & Purity \\
    \hline
    \hline
    \lnlns & 10.6\% & 60-80\% & 90\% \\
    \qqlns & 43.8\% & 70-85\% & 90\% \\
    \qqqqs & 45.6\% & 80-90\% & 80\% \\
    \hline
  \end{tabular}
  \label{tab:wwsel}
\end{table}

The golden channel for most LEP W analyses are the semi-leptonic
\qqlns\ decays, as these can be selected with high efficiency
and very low backgrounds in all but the \qqtn\ channel.
The presence of the lepton also allows the charge of the W boson
to be tagged.
The fully hadronic \qqqqs\ decays are also useful, although there
are significant backgrounds from two-fermion \Zqq\ events, and
with four jets there is an ambiguity in pairing the observed jets
to the underlying W bosons.
The fully leptonic \lnlns\ events are less useful due to their
low rate and the large missing momentum carried by the two 
final state neutrinos.

\begin{figure}[ht]
  \begin{center}
    \epsfig{file=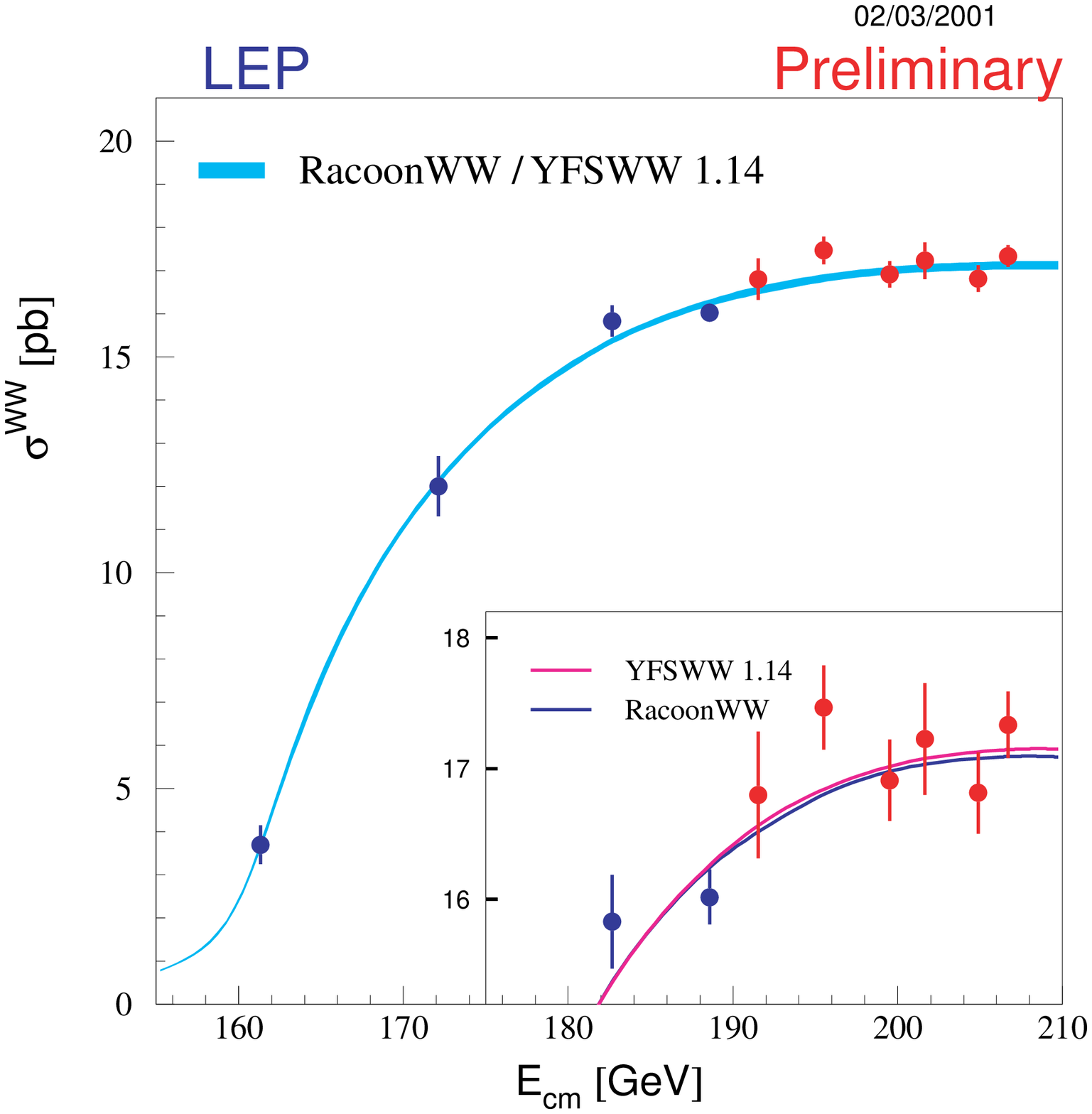,width=8cm}
    \caption{\it
      The \CC\ production cross section vs. $\protect\roots$.
      \label{fig:wwxsec}
      }
 \end{center}
\end{figure}

Figure~\ref{fig:wwxsec} shows the measured \WW\ production cross
section from threshold up to the highest collision energy reached 
at \LEPII\cite{bib:lepxsec}.
These data are compared with two recent theoretical calculations
which include a more complete treatment of ${\cal O}(\alpha)$ 
radiative corrections through the double pole 
approximation\cite{bib:YFSWW,bib:RAC}.
Good agreement is seen between the data and the newer calculations,
which predict an overall rate (2.3--2.4)\% lower than the older
\GENTLE\cite{bib:GENTLE} semi-analytic calculation.
\subsection{W-decay branching fractions at \LEPII}
Using the production rates observed in the ten unique \WW\
decay topologies, each LEP collaboration performs fits to
its own data to extract the leptonic branching fractions
individually, and with the additional constraint of lepton 
universality fits are performed for the hadronic branching 
fraction as well.
In both cases, it is assumed that $\Br(\Wtoqq)+\Br(\Wtoln)=1$.
The combined LEP results are shown in 
Table~\ref{tab:brw}\cite{bib:lepxsec}.

\begin{table}[t]
  \centering
  \caption{\it LEP combined W branching fractions.
    }
  \vskip 0.1 in
  \begin{tabular}{|l|c|c|c|} \hline
    Decay & Observed & SM Expectation \\
    \hline
    \hline
    $\Br(\Wtoen)$ & $10.54\pm0.17$\% & 10.83\% \\
    $\Br(\Wtomn)$ & $10.54\pm0.16$\% & 10.83\% \\
    $\Br(\Wtotn)$ & $11.09\pm0.22$\% & 10.83\% \\
    \hline
    $\Br(\Wtoqq)$ & $67.92\pm0.27$\% & 67.51\% \\
    \hline
  \end{tabular}
  \label{tab:brw}
\end{table}

The hadronic branching fraction can be interpreted as a 
measurement of the sum of the squares of the six elements 
of the CKM mixing matrix, \Vij, which do not involve the 
top quark:
\begin{equation}
     {{\Br(\Wtoqq)}\over{1-\Br(\Wtoqq)}} =
 \left( 1+\frac{\alpha_s(\Mw)}{\pi} \right)
\sum_{i={\mathrm{u,c}}; \, j={\mathrm{d,s,b}}} \Vij^2.
\end{equation}
Taking $\alpha_s(\Mw)$ to be $0.120\pm0.005$,
the branching fraction $\Br(\Wtoqq)$ from the combined LEP data 
yields
\begin{equation}
 \sum_{i={\mathrm{u,c}};\, j={\mathrm{d,s,b}}}
  \Vij^2 = 2.039\pm0.025,
\end{equation}
which is consistent with the value of 2 expected from unitarity in
a three-generation CKM matrix.

Using the experimental knowledge of the sum,
$\Vud^2+\Vus^2+\Vub^2+\Vcd^2+\Vcb^2 = 1.048\pm0.007$\cite{bib:pdg}, 
the above result can be interpreted as a measure of \Vcs\ which is the 
least well determined of these matrix elements:
\begin{equation}
  \Vcs = 0.996\pm0.013.
  \label{eqn:vcs}
\end{equation}

A more direct determination of $\Vcs$ is also performed by the
LEP collaborations by counting the number of charm
quarks produced in W events\cite{bib:rcw}.
For example, Opal uses a multivariate analysis similar to what
is used to tag b-quarks at \LEPI\ to extract a value of
\begin{equation}
\Rcw = 0.481\pm0.042\mathrm{(stat.)}\pm0.032\mathrm{(syst.)}.
\end{equation}
Using the relation
\begin{equation}
\Gamma(\mathrm{W}\rightarrow\mathrm{cX}) = 
\frac{C G_F \Mw^3}{6 \sqrt{2} \pi}(\Vcd^2+\Vcs^2+\Vcb^2),
\end{equation}
this measurement can again be interpreted in terms of \Vcs,
with fewer assumptions than the value shown in Equation~\ref{eqn:vcs}:
\begin{equation}
  \Vcs = 0.969\pm0.058.
\end{equation}
\subsection{W boson production at the Tevatron}
\begin{figure}[ht]
  \begin{center}
    \epsfig{file=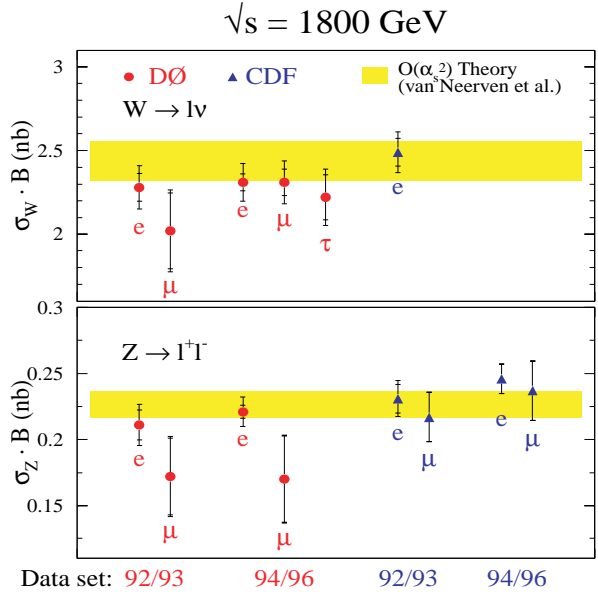,width=8cm}
    \caption{\it
      Tevatron production rates of W and Z bosons.
      \label{fig:tevbr}
      }
 \end{center}
\end{figure}

At the Tevatron, measurements are made of the production cross section
times leptonic branching fraction for both W and Z bosons.
These results from \RunI\ are shown in Figure~\ref{fig:tevbr}.
Due to a different luminosity convention, the cross section
quoted by CDF is 6\% higher than that quoted by D0.
The experimental uncertainties on these measurements are typically
around 2\% with a theoretical uncertainty of typically 3\%,
dominated by the knowledge of the parton distribution functions (PDFs).
Since the luminosity uncertainty at the Tevatron is comparable,
around 4\%, at \RunII\ it may well be advantageous to use the
measurement of $\Br(\Wtoen)$ as a measure of the delivered 
luminosity.

Using the ratio of electron production by W and Z bosons,
the Tevatron collaborations perform a rather accurate indirect 
measurement of the W width.
This ratio can be written as
\begin{equation}
\frac{\sigma\cdot\Br(\Wtoen)}{\sigma\cdot\Br(\Ztoee)} = 
\frac{\sigW}{\sigZ}\frac{\Gz}{\Gamma(\Ztoee)}
\frac{\Gamma(\Wtoen)}{\Gw},
\end{equation}
where the ratio $\sigW/\sigZ$ is known from QCD\cite{bib:qcdratio}, 
and the \LEPI\ measurement is used for $\Br(\Ztoee)$\cite{bib:ztoee}.
This can then be solved to extract the W branching
fraction, which for CDF and D0 combined is
\begin{equation}
\Br(\Wtoen) = 10.45\pm0.20 \, \%.
\end{equation}
If it is additionally assumed that the electronic width
of the W boson is equal to the Standard Model prediction 
$\Gamma(\Wtoen) = 226.5\pm0.3$~MeV\cite{bib:wtoen}, this can be
interpreted as an indirect measurement of the total
width,
\begin{equation}
\Gw(\mathrm{CDF+D0}) = 2.167\pm0.040 \, \mathrm{GeV},
\end{equation}
in good agreement with the Standard Model
prediction of $\Gw(\mathrm{SM}) = 2.093\pm0.003$~GeV.
%
%
\section{Electroweak Couplings}
\label{sec:tgc}
One of the fundamental predictions of the electroweak theory
is the self-coupling of the gauge bosons due to the non-Abelian
nature of the gauge fields.
In general, there are seven unique couplings for each $WWV$
vertex which are Lorentz invariant, where $V$ denotes either
of the neutral vector bosons.
From an experimental point of view, this represents far too
many individual parameters to measure, and several assumptions
and constraints have been applied to reduce this to a more
manageable set of three.
First, considerations of gauge invariance, $C$, $P$, and $CP$ 
conservation are used to reduce this set down to five, 
which are usually denoted as $\kg,\kz,\lag,\laz,\gz$.
The Standard Model predicts that only three of these couplings
are non-zero, with $\kg=\kz=\gz=1$.
More familiar static properties of the W boson are then some
function of the general couplings, for example the magnetic
dipole moment is given as $\mu_\mathrm{W}=e(1+\kg+\lag)/2\Mw$.

The precision electroweak data collected at \LEPI\ and the
SLC can be used to further constrain the allowed couplings
down to a set of three.
This precision data implies the relations $\dkz = -\dkg\tanw+\dgz$
and $\lag=\laz$, where $\dkz=\kz-1$ is the deviation from the
Standard Model prediction.
\subsection{Triple gauge couplings at \LEPII}
\begin{figure}[ht]
  \begin{center}
    \epsfig{file=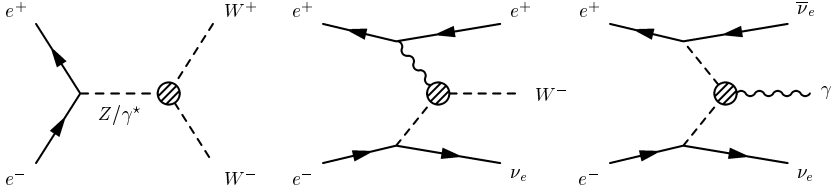,width=11cm}
    \caption{\it
      \LEPII\ processes sensitive to triple gauge couplings.
      \label{fig:tgcdiag}
      }
 \end{center}
\end{figure}

At \LEPII, triple gauge coupling (TGC) analyses consider
the processes shown in Figure~\ref{fig:tgcdiag}.
Most of the sensitivity arises from W-pair production,
although the single-W diagram and single photon
production help constrain $\kg$ which is otherwise poorly
measured in high energy \epem\ collisions.
The presence of anomalous TGCs will modify both the total 
cross section and the differential cross sections of these
processes, and the LEP analyses make use of all of this 
information to extract TGC limits.

\begin{figure}
  \begin{center}
    \epsfig{file=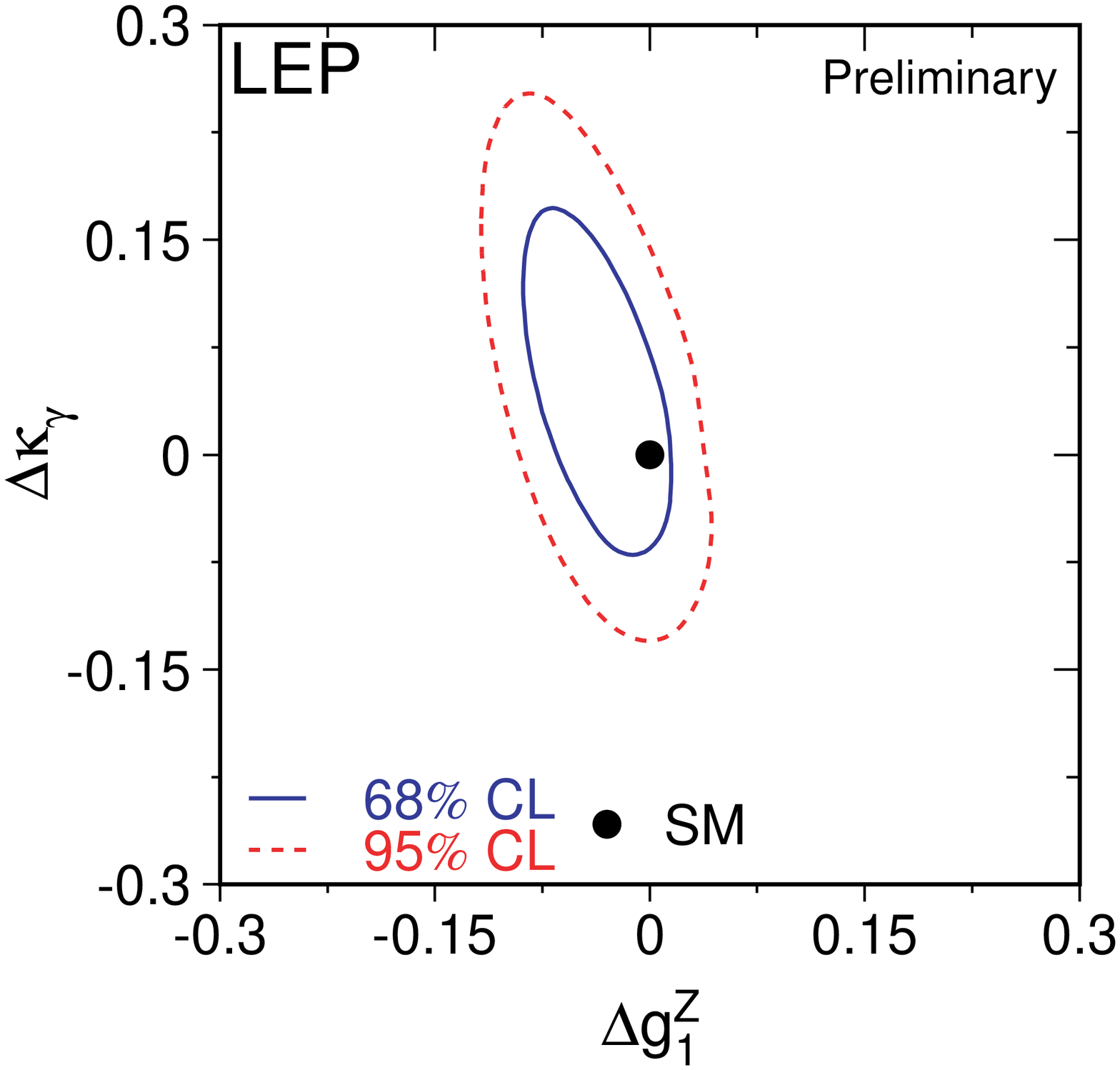,width=5.25cm}
    \hspace{0.5cm}
    \epsfig{file=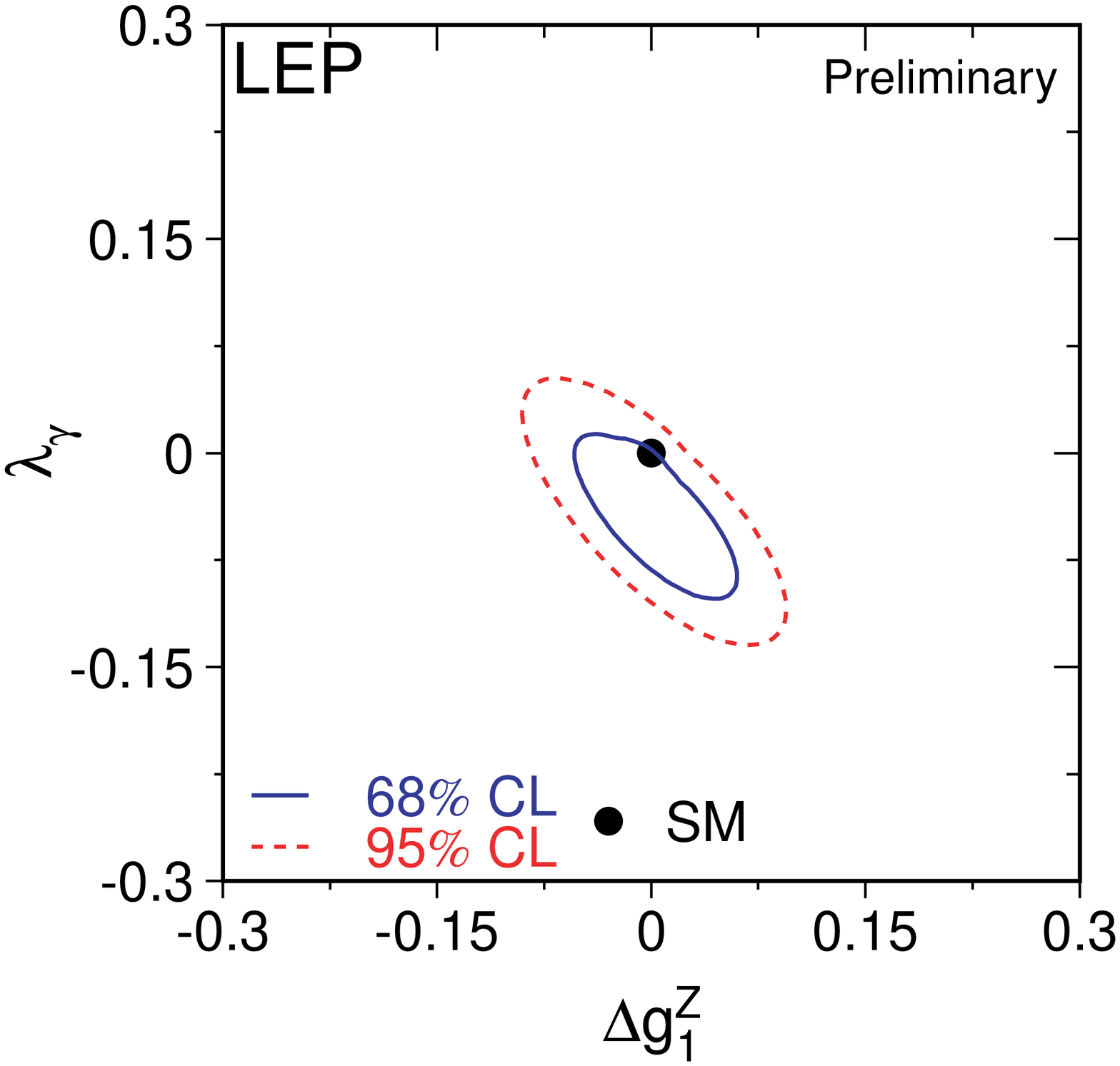,width=5.25cm}
    \caption{\it
      \LEPII\ combined two-dimensional TGC fits.
      \label{fig:tgcfits}
      }
 \end{center}
\end{figure}

LEP combined limits are obtained for both single-parameter
fits with the other two anomalous couplings fixed to zero,
and for multi-dimensional fits where two or all three
anomalous couplings are allowed to deviate from zero.
The most recent results for the LEP single parameter fits
are
\begin{eqnarray}
\dkg &=& -0.002 \pm 0.066 \nonumber \\
\dgz &=& -0.025 \pm 0.026 \\
\lag &=& -0.036 \pm 0.028 \nonumber
\end{eqnarray}
where the errors indicate the one sigma (68\% CL) errors of
the fit.
Examples of multi-dimensional confidence level regions are
shown in Figure~\ref{fig:tgcfits}.
These results date from ICHEP~2000\cite{bib:i00_tgc} and
do not include potentially large corrections arising from
${\cal O}(\alpha)$ radiative corrections which have recently
been included into standard Monte Carlo generators.
New combined results taking these effects into account
should be available by the Summer conferences of 2001.
At any rate, the precision of these measurements is at the
3\% level for all but \dkg, which is the most difficult
coupling to measure at \LEPII.

\subsection{Triple gauge couplings at the Tevatron}
The Tevatron collaborations can also measure triple gauge
couplings through either associated production of the
W boson accompanied by a neutral gauge boson, or through
direct W-pair production.
The most useful final state for associated production
is \lnplusg, where a real photon is radiated directly
from the W.
Equivalent diagrams involving a Z boson are marginal at
\RunI, but should become useful at \RunII.
In the case of W-pair production, the \lnlns\ and \qqlns\ final
states are analyzed.
Similarly to LEP, the sensitivity to TGCs at the Tevatron
arise from both an increase in the total rate for these
processes, but also from the shape of the various differential
cross sections.
D0, for example, performs a combined analysis to the 
\lnplusg, \lnlns, and \qqlns\ channels using both the
overall rate and the \ptw\ distributions as fit
variables.

Fitting for two TGC parameters, D0 obtains the following 95\% 
confidence level limits
\begin{equation}
\begin{array}{rcccl}
-0.25 & \leq & \Delta\kappa & \leq & +0.31 \\
-0.18 & \leq & \lambda & \leq & +0.18
\end{array}
\end{equation}
where in each case the second coupling is fixed to be zero\cite{bib:d0tgc}.
With the increased luminosity and collision energy
expected at \RunII, these results should improve by a factor
of two to three, which will make D0 competitive with any
single LEP experiment.
\subsection{Quartic gauge couplings}
\begin{figure}[ht]
  \begin{center}
    \epsfig{file=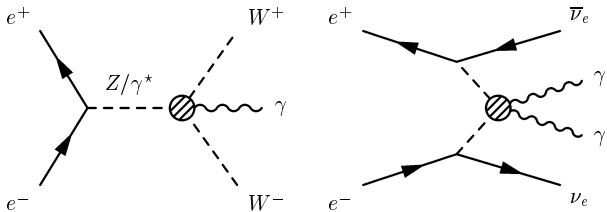,width=8cm}
    \caption{\it
      \LEPII\ processes sensitive to anomalous quartic gauge couplings.
      \label{fig:qgcdiag}
      }
 \end{center}
\end{figure}

The Standard Model also predicts a genuine four-point 
interaction involving only gauge bosons.
Some of the LEP collaborations have performed analyses
using the diagrams shown in Figure~\ref{fig:qgcdiag}
to attempt to observe these quartic gauge couplings (QGCs).
Since the predicted strength of this coupling in the
Standard Model is negligible at \LEPII, any deviation
from zero would be evidence for anomalous QGCs.

The Lorentz structure of the four-boson vertex can be 
parameterized in a manner similar to that used in the TGC case.
The LEP collaborations have explored the possibility
of two $CP$-conserving parameters $a_0$ and $a_c$ which
involve the \WWgg\ vertex\cite{bib:BandB},
along with a single $CP$-violating coupling $a_n$ which
involves the \WWZg\ vertex\cite{bib:SandW}.
In addition, analyses have been performed for QGCs involving
purely neutral gauge bosons.

\begin{figure}
  \begin{center}
    \epsfig{file=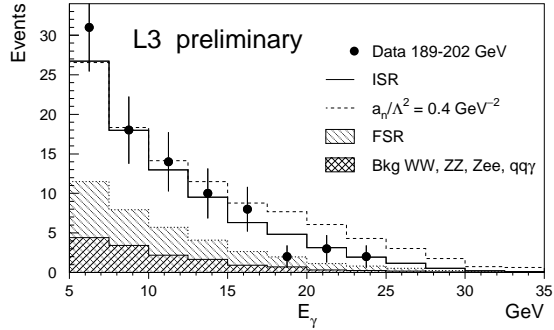,width=8cm}
    \caption{\it
      L3 photon spectrum for \WWg\ QGC analysis.
      \label{fig:L3qgc}
      }
 \end{center}
\end{figure}

In the \WWg\ final state, the signature of anomalous QGCs
is an enhancement of the hard photon spectrum above that
expected from other radiative processes.
An example of this spectrum from L3 is shown in 
Figure~\ref{fig:L3qgc}.

An unofficial LEP combination of the available QGC results 
from Aleph, Opal, and L3 was presented at 
Moriond 2001\cite{bib:m01-qgc}.
The combined limits at the 95\% CL in units of $\mathrm{GeV}^2$ are
\begin{equation}
\begin{array}{rcccl}
-0.022 & \leq & a_0/\Lambda^2 & \leq & +0.021 \\ 
-0.043 & \leq & a_c/\Lambda^2 & \leq & +0.058 \\
-0.22  & \leq & a_n/\Lambda^2 & \leq & +0.20. 
\end{array}
\end{equation}
%
%
\section{W Mass and Width}
\label{sec:mass}
When combined with other precision electroweak data, most
notably the mass of the Z boson, the W boson mass becomes 
a fundamental prediction of the Standard Model.
Testing this prediction with ever increasing precision
has been a goal of particle physics over the
last decade, both to verify the Standard Model, but
also to gain insight into what may lie beyond.
In this section, the measurements of \Mw\ at the Tevatron
and \LEPII\ are described.
\subsection{W mass at the Tevatron}
\begin{figure}[ht]
  \begin{center}
    \epsfig{file=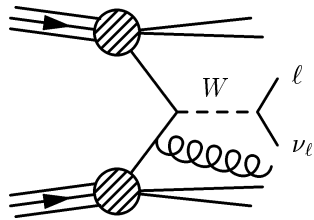}
    \caption{\it
      W boson production in \ppbar\ collisions.
      \label{fig:qqwprod}
      }
 \end{center}
\end{figure}

It is customary to think of W production at the Tevatron
in terms of the simple process \qqWln, while in reality
the actual event is considerably more messy.
As shown in Figure~\ref{fig:qqwprod}, along with the W boson
comes hard gluon radiation which must balance the W boson \pt,
along with the ever-present spectator and multiple
interaction background from the \ppbar\ collision.

Since there is essentially no information available regarding
the longitudinal boost of the W boson, CDF and D0 both use
the transverse mass as the fundamental observable in their
W mass analysis.
The transverse mass is given by
\begin{equation}
\mt = \sqrt{2 \Etl \Etnu (1-\cos\phi_{\ell\nu})},
\end{equation}
where the neutrino is defined in terms of the missing transverse
momentum $\vecptnu = -(\vecptl + \vecU)$ with \vecU\ representing 
the transverse momentum of the hard QCD recoil.

\begin{figure}
  \begin{center}
    \epsfig{file=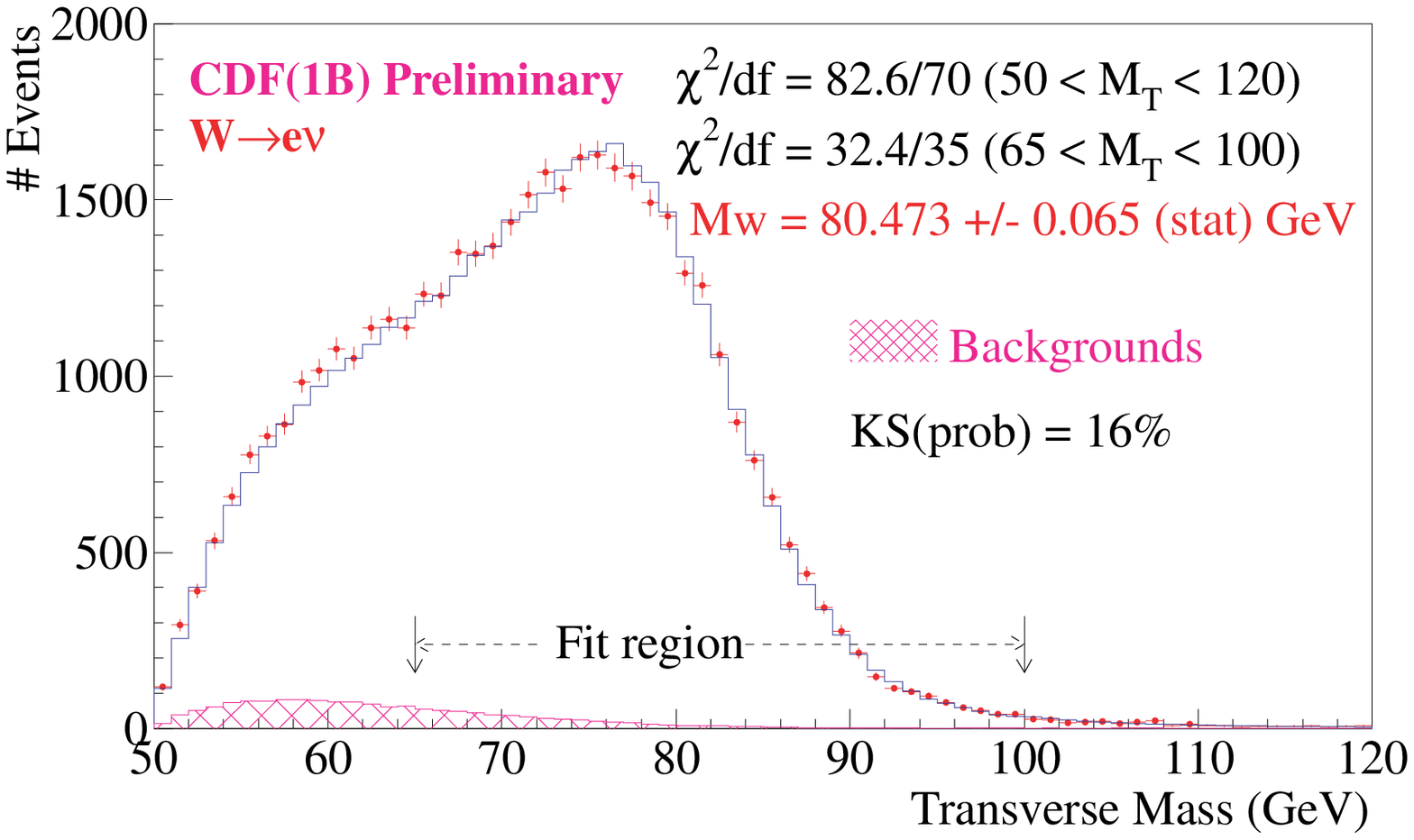,width=8cm}
    \caption{\it
      CDF transverse mass spectrum for \Wtoen.
      \label{fig:cdfmass}
      }
 \end{center}
\end{figure}

In general terms, the mass analysis at the Tevatron involves
fitting the transverse mass spectrum observed in the data
with a Monte Carlo prediction of this spectrum which depends
upon \Mw.
An example of such a fit from CDF is shown in 
Figure~\ref{fig:cdfmass}.
There are a number of individual components necessary to
construct the model of the \mt\ spectrum, each of which
contributes to the overall systematic error of the measurement.
The three most important are described below.

Firstly, the energy scale and resolution of the detector to
leptons and hadrons must be precisely modelled.
This modelling can be constrained using control samples
found in the data, most notably \Zll\ decays which have
a very similar topology to the W decays of interest.
D0, for example, calibrates their detector response to
electrons by fitting to the observed invariant mass 
distributions for \pieeee, \Jee, and \Zee.
The uncertainty on the detector response is largely
driven by the statistics available in the available 
control samples, and will improve considerably in \RunII.
A list of typical uncertainties\footnote{In this section, `typical' 
usually means CDF.} is shown in 
Table~\ref{tab:cdfsyst}\cite{bib:m01-dg}.

\begin{table}[ht]
  \centering
  \caption{\it Typical Tevatron W mass uncertainties per channel.
        An indication is also given whether an uncertainty is
	largely statistical or correlated between CDF and D0.
    }
  \vskip 0.1 in
  \begin{tabular}{|l|c|cc|} \hline
    Source & Uncertainty (MeV) & Statistical? & Correlated? \\ 
    \hline
    \hline
    statistical   & 60 - 100 & yes &  \\
    \hline
    scale         & 80 & yes &     \\
    resolution    & 25 & yes &     \\
    backgrounds   & 15 &     &     \\
    recoil        & 40 & yes &     \\
    PDFs          & 15 & yes & yes \\
    higher orders & 15 &     & yes \\
    \hline
  \end{tabular}
  \label{tab:cdfsyst}
\end{table}

Secondly, the transverse momentum spectrum of the QCD recoil
must be modelled properly.
Again, control samples in the data provide powerful constraints
to tune the production models used in the Monte Carlo generators.
By comparing the transverse momentum observed in \Zll\ events,
where there is no missing neutrino, adequate statistics can
be found to limit this uncertainty to an acceptable level.

Finally, the parton density functions (PDFs) used to describe
the constituents of the incoming \ppbar\ system must be constructed.
Knowledge of these PDFs comes from a wide variety of data
made at the Tevatron and elsewhere\cite{bib:cteq}.
Although this is currently a rather modest uncertainty
at the level of 15~MeV, since it is correlated between
both experiments and rather difficult to constrain directly
with Tevatron data, it could become a limiting systematic
at \RunII.

While the general analysis strategy for the W mass is similar 
between CDF and D0, there are significant differences in the details.
CDF analyzes \Wtoen\ and \Wtomn\ events where the lepton
is reconstructed in the central detector region $(|\eta_\ell|<1)$.
D0 uses only \Wtoen, although electrons in the forward detector
region are also included $(1.5<|\eta_\ell|<2.5)$.
In addition, D0 performs a fit not only to the transverse mass
spectrum, but also to the transverse momentum spectra \pte\ and \ptnu.

The final W mass results from the Tevatron collaborations
for \RunI\ are
\begin{equation}
\begin{array}{lccl}
\Mw(\mathrm{CDF}) &=&80.433\pm0.079&\mathrm{GeV} \\
\Mw(\mathrm{D0})  &=&80.482\pm0.091&\mathrm{GeV},
\end{array}
\end{equation}
which when combined accounting for correlated systematic 
uncertainties yields a W mass measurement of\cite{bib:tevmass}
\begin{equation}
\Mw(\ppbar) = 80.452\pm0.062 \, \mathrm{GeV}.
\end{equation}
It should be noted that D0 is updating their `final' \RunI\
result in Summer 2001 by adding additional \Wtoen\ events 
where the electrons went into detector regions previously 
considered unusable.
With this additional data, the combined D0 results
should be quite competitive with the final CDF \RunI\ measurement.

The prospects for the W mass measurement are quite encouraging.
With 2~\fbi\ expected in \RunIIa, there is significant room
for improvement both in the statistical and systematic uncertainties,
since many of these systematics are limited by available control
samples.
It is expected that a total uncertainty of under 40~MeV per
experiment can be achieved in \RunIIa.
\subsection{W mass at \LEPII}
While the sample of W bosons collected at \LEPII\ is significantly
smaller that that collected at the Tevatron, the benefits of a
clean \epem\ collision environment allow more information to be
extracted from each event recorded.
In particular, the well defined initial state allows the use of
constrained kinematic fitting to greatly improve the mass
resolution on the reconstructed W boson.
All four LEP collaborations measure the W mass by a method of
direct reconstruction based on constrained kinematic fitting, 
although the exact details of each analysis can vary significantly.

Four constraints are imposed on the kinematics of each event
requiring that the initial \epem\ system is at rest in the center
of mass and has an energy equal to twice the incoming beam energy.
In some analyses, an additional fifth constraint forcing the two
W boson masses to be equal, is also imposed.
For the semi-leptonic \qqlns\ final states, there is a single
neutrino which is not measured, leaving a two constraint fit
after the equal mass constraint is applied.
For the fully hadronic \qqqqs\ final states, there is no
missing momentum and a full four or five constraint fit is
possible depending upon whether the equal mass constraint is used.

\begin{figure}
  \begin{center}
    \epsfig{file=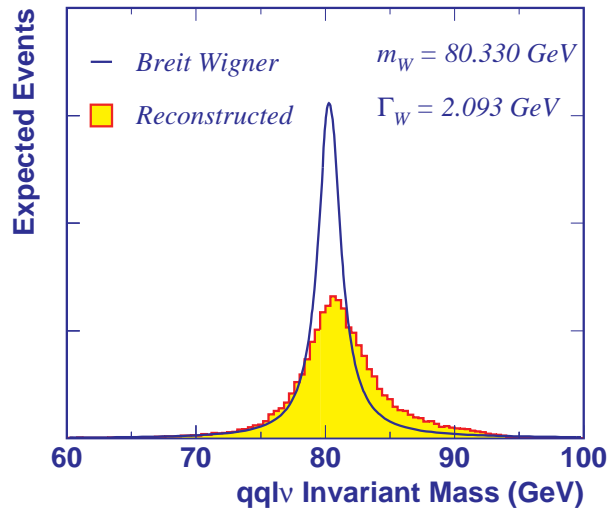,width=8cm}
    \caption{\it
      Reconstructed \qqlns\ invariant mass spectrum from Monte Carlo.
      \label{fig:bwdist}
      }
 \end{center}
\end{figure}

The reconstructed mass spectrum, as shown for example in 
Figure~\ref{fig:bwdist}, looks rather different than a pure
relativistic Breit--Wigner distribution for several reasons.
Firstly, the presence of initial state radiation (ISR) means
that the actual energy producing the W pairs is always less than
twice the incoming beam energy.
This tends to cause a tail towards higher invariant masses, as
the collision energy assumed in the kinematic fit is overestimated.
Secondly, effects of detector resolution tend to significantly
smear out the lineshape as the experimental resolution is not
significantly better than the W boson width for most channels.
Finally, the effects of non-\CC\ diagrams, particularly in the
\qqen\ channel, distort the lineshape due to the interference
with the resonant production amplitudes.

As a result, the W boson mass can not be extracted by simply 
fitting an analytic Breit--Wigner shape, but rather must rely
on a Monte Carlo simulation of these various effects to model
the dependence of the spectrum on \Mw.
Delphi uses a likelihood convolution technique to construct
a likelihood $\cal{L}(\Mw)$ for each event.
The remaining three collaborations currently use
re-weighting techniques to construct Monte Carlo templates
at arbitrary \Mw\ to be fit to the data 
distributions.

The fully hadronic \qqqqs\ final state has the additional
complication of a three-fold ambiguity in pairing the observed
jets to the underlying W bosons.
Aleph and Opal use an additional jet pairing algorithm to choose
for each event which pairing is most likely to be correct.
L3 uses the single pairing which yields the highest kinematic
fit probability, while Delphi can naturally use all three 
combinations appropriately weighted in the likelihood convolution
technique.

\begin{figure}[ht]
  \begin{center}
    \epsfig{file=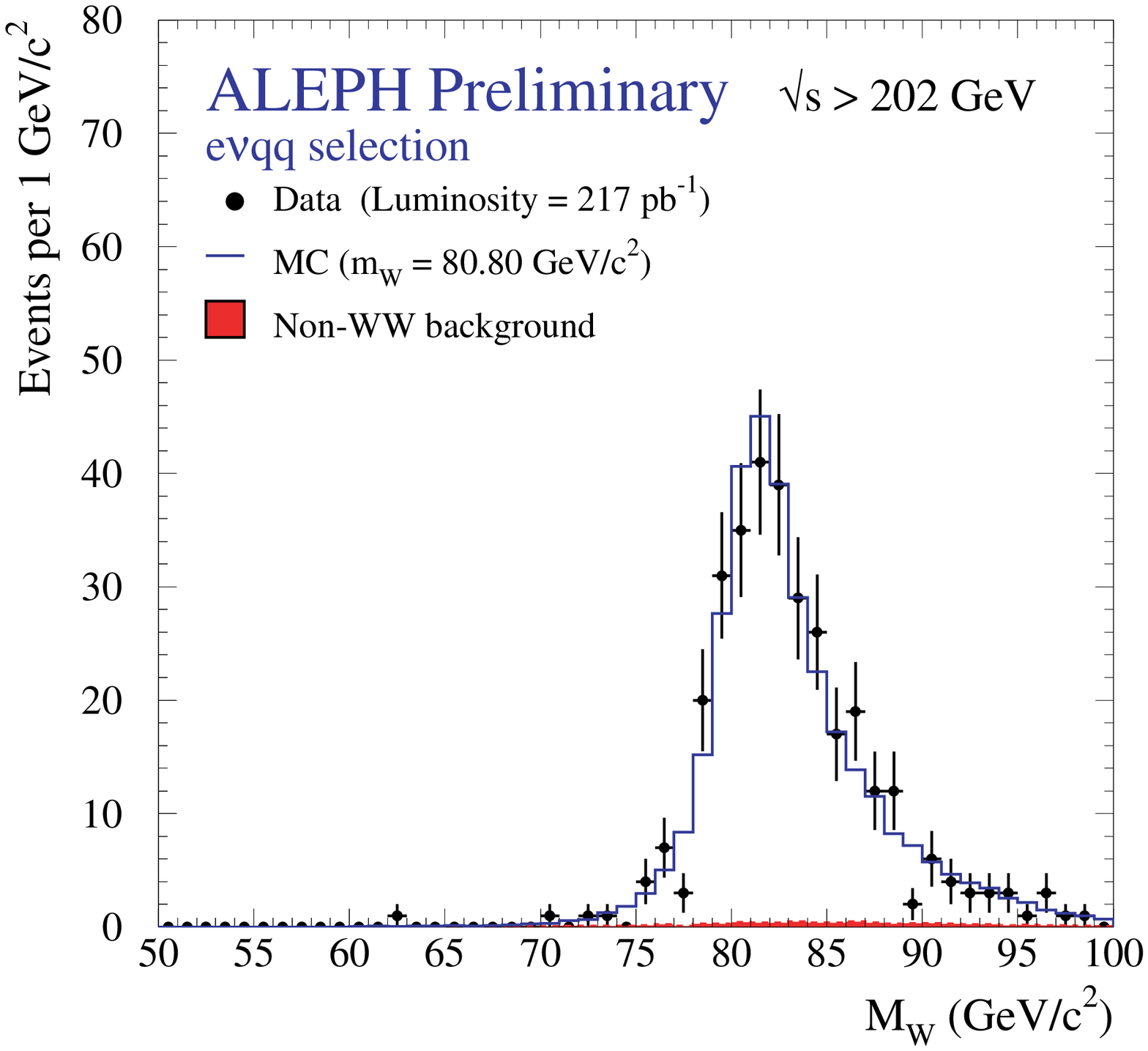,width=4.5cm}
    \epsfig{file=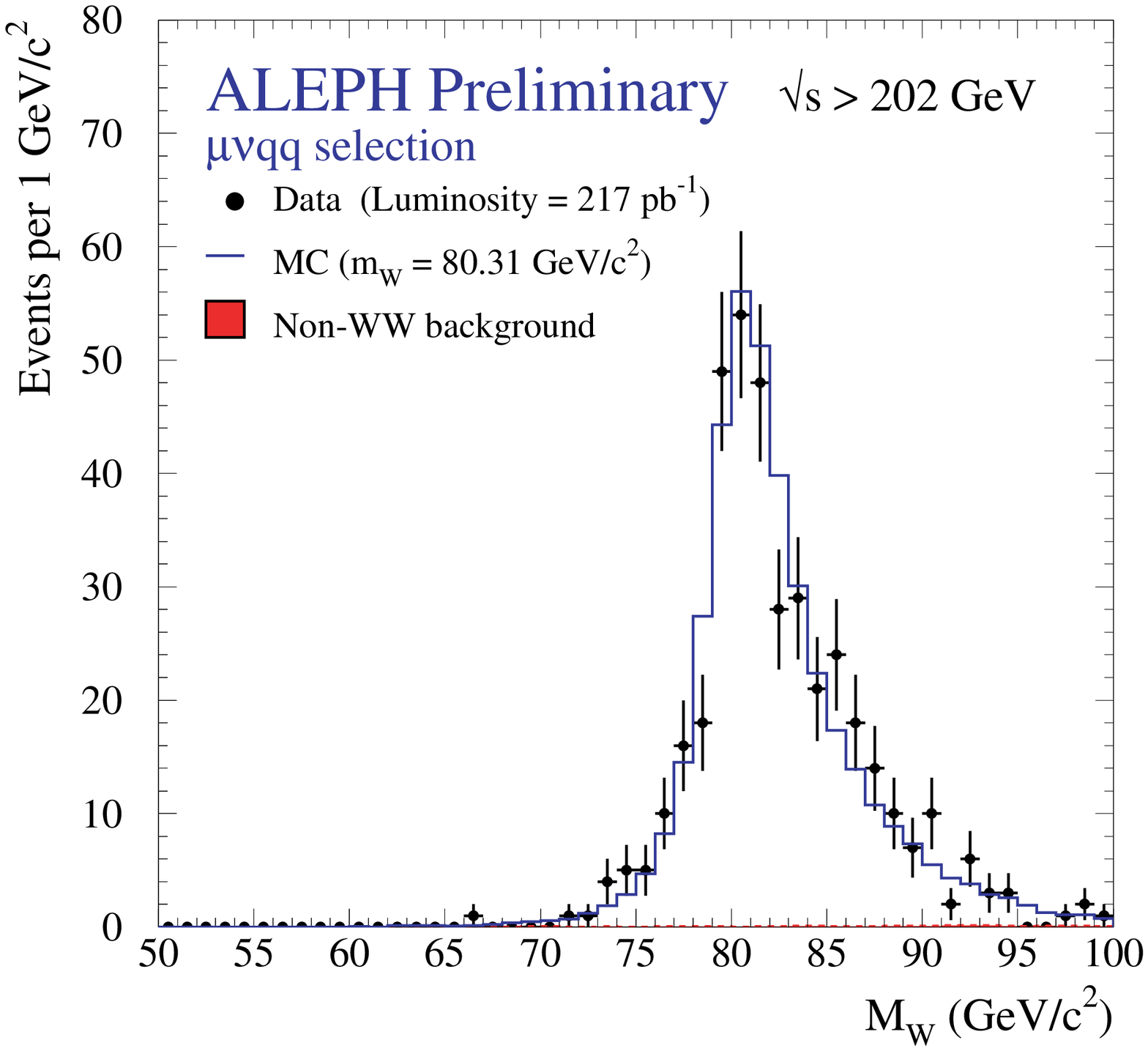,width=4.5cm} \\
    \epsfig{file=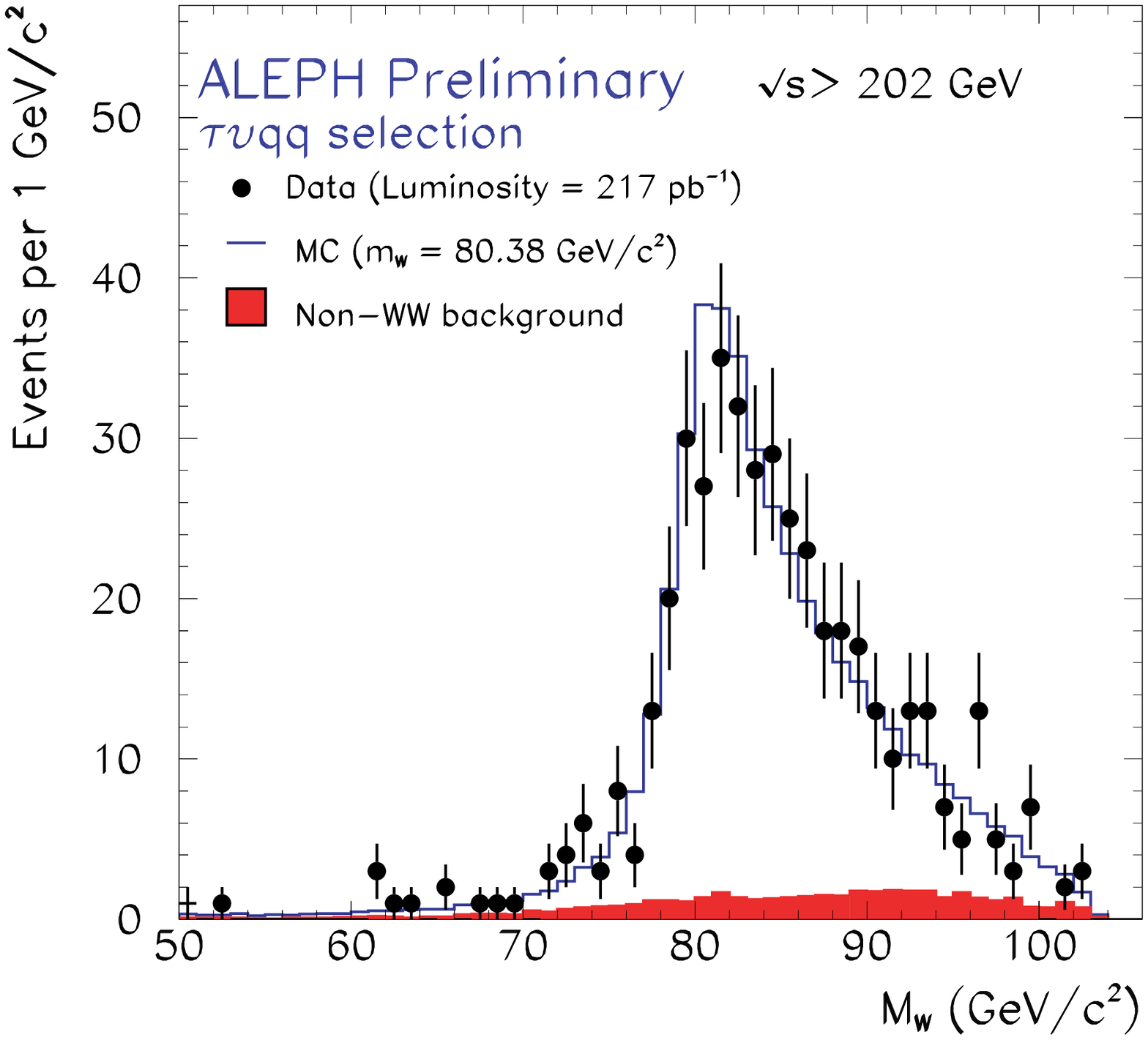,width=4.5cm}
    \epsfig{file=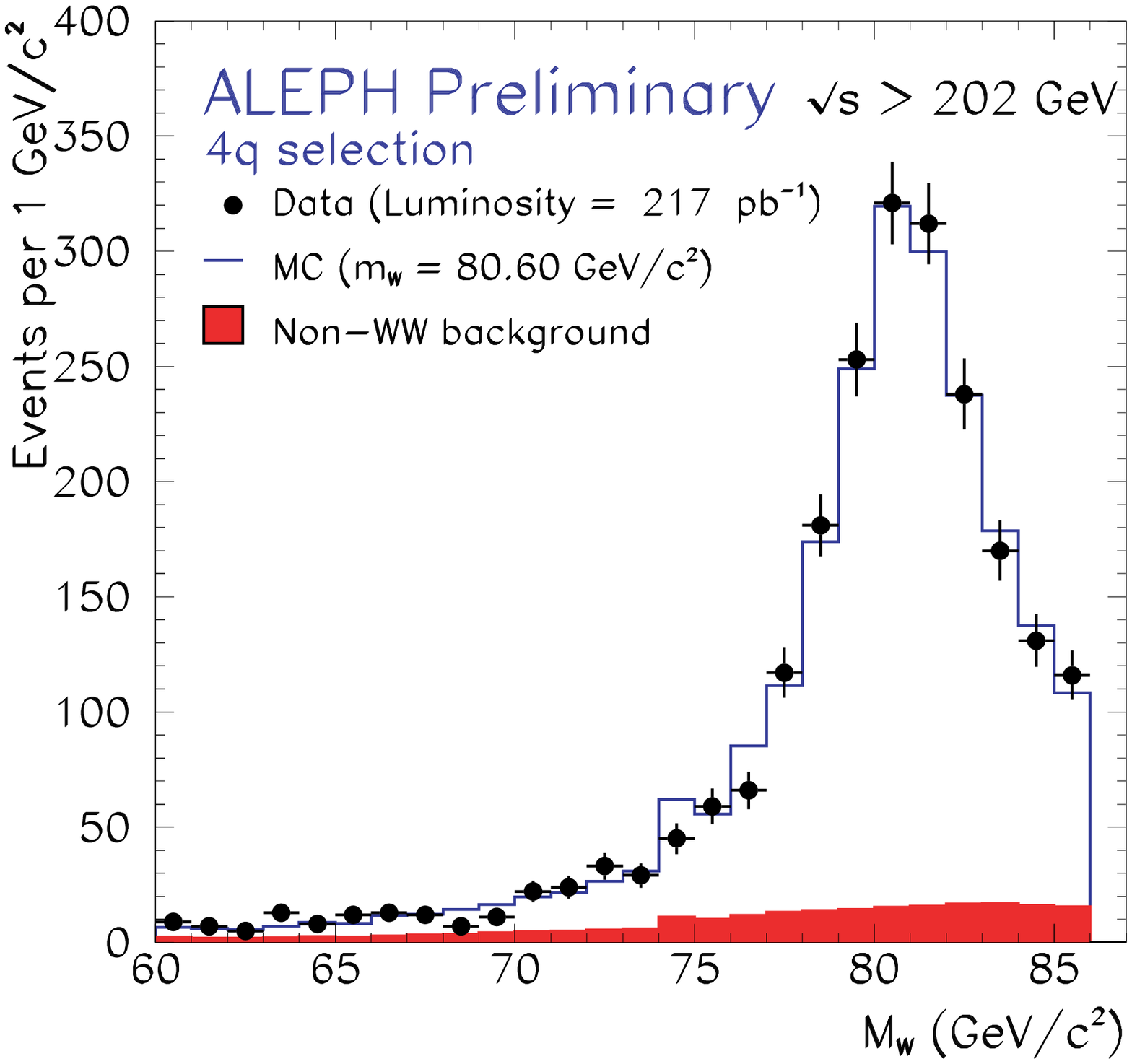,width=4.5cm}
    \caption{\it
      Reconstructed mass spectra from the Aleph collaboration.
      \label{fig:aleph-mw}
      }
 \end{center}
\end{figure}

Figure~\ref{fig:aleph-mw} shows some mass spectra compared to
the Monte Carlo fits for Aleph data recorded during the final
year of LEP running.

The four LEP collaborations combine their results taking into
account systematic uncertainties which are correlated between
channels, experiments, and years of LEP running.
This combination procedure is still evolving, as better
information about the nature of these various correlations
becomes available.
The preliminary combined LEP W mass result from direct reconstruction
presented at Moriond 2001 is
\begin{equation}
\Mw(\mathrm{LEP}) = 80.447\pm0.026\stat\pm0.030\syst\,\mathrm{GeV},
\end{equation}
where the component of the total uncertainty coming from statistical
and systematic sources is indicated\cite{bib:m01-lepmw}.
As can be seen in the detailed breakdown of these uncertainties shown
in Table~\ref{tab:lepsyst}, the \qqqqs\ channel has rather large
uncertainties associated with Bose--Einstein correlations and color
reconnection.
Due to these uncertainties, the \qqqqs\ channel carries a weight
of only 27\% in the combined result, even though it is statistically
more precise.
If there were no sources of systematic uncertainty, the combined 
statistical error would be 22~MeV.

\begin{table}[ht]
  \centering
  \caption{\it 
	Uncertainties on the LEP combined W mass.
    }
  \vskip 0.1 in
  \begin{tabular}{|l|c|c|c|} \hline
    Uncertainty & \qqlns\ (MeV) & \qqqqs\ (MeV) & Combined (MeV) \\
    \hline
    \hline
    ISR/FSR       &  8 &  8 &  7 \\
    Hadronization & 19 & 17 & 18 \\
    Detector      & 11 &  8 & 10 \\
    Beam Energy   & 17 & 17 & 17 \\
    Color Recon.  &  - & 40 & 11 \\
    Bose-Einstein &  - & 25 &  7 \\
    Other         &  4 &  5 &  3 \\
    \hline
    Total Syst.   & 29 & 54 & 30 \\
    \hline
    Statistical   & 33 & 31 & 26 \\
    \hline\hline
    Total         & 44 & 62 & 40 \\
    \hline
  \end{tabular}
  \label{tab:lepsyst}
\end{table}

The most important uncertainties for the LEP W mass measurement are
related to the modelling of the detector response, modelling
the hadronization of quarks into jets, understanding the LEP beam
energy, and the final state interactions mentioned before.
To limit the uncertainty from detector modelling, the LEP collider
was run from time to time throughout each running year at the
\Zz\ resonance to take large samples of \Zll\ and \Zqq events.
These samples, as well as similar samples at high energy, are then
used to study the detector response to leptons and jets and limit 
the deficiencies in the detector models.
Hadronization uncertainties are estimated by comparing different
Monte Carlo implementations of the hadronization process, including
\JETSET\cite{bib:PYTHIA}, \HERWIG\cite{bib:HERWIG}, 
and \ARIADNE\cite{bib:ARIADNE}.
Knowledge of the LEP beam energy is critically important as it sets
the overall energy scale for the W mass measurement, and any uncertainty
in is completely correlated between all four LEP collaborations.
More detailed information about the beam energy measurement at \LEPII\
can be found elsewhere\cite{bib:lepenergy}.

The broad issue of final state interactions (FSI) involves any effect in
the \qqqqs\ channel by which the decay products of the two W bosons
interact with each other.
Since standard Monte Carlo models assume that these two systems decay
independently, these interactions can exchange momentum between the
two W systems and distort the final W lineshape.
Two specific phenomena which are known to exist, but with rather
uncertain strengths, are color reconnection (CR) and Bose--Einstein
correlations (BEC).

Color reconnection involves the rearrangement of the color strings
in the hadronizing system, essentially gluon exchange, in the
non-perturbative phase of the hadronization process.
Only a few models exist to describe this process, and these 
tend to have limited predictive power\cite{bib:CR}.
Bose--Einstein correlations between identical particles are 
well established in hadronic \Zz\ decays at \LEPI.
Since the W boson decay length (0.1~fm) is significantly shorter
than the hadronization scale (1~fm), it is entirely plausible
that there can be additional BE effects between particles originating
from different W bosons in \qqqqs\ events.
Because of the way Monte Carlo generators operate, however, modelling
this purely quantum mechanical effect is extremely difficult, and
imperfect knowledge of the space-time evolution of the hadronizing
system adds to the difficulty\cite{bib:BE}.

Overall, the strategy of the LEP experiments to address these difficult
questions is to use the various models as a guide for the types of
effects which can be expected, but to constrain the allowed magnitude
of any FSI effect using the data directly.
One way to broadly check that large FSI effects are not
present is to compare the W mass measured in the \qqqqs\ channel to
that measured in the \qqlns\ channel.
Including all uncertainties with their correlations aside from those
related to FSI directly, the LEP collaborations find
\begin{equation}
\Mw(\qqqqs) - \Mw(\qqlns) = +18\pm46\,\mathrm{MeV},
\end{equation}
which is consistent with no final state interactions.

Color reconnection effects tend to enhance or suppress particle
production in the regions between the main jets.
Currently, all four LEP collaborations are pursuing analyses
aimed at measuring the particle flow distribution in \qqqqs\
final states with the ultimate aim of discriminating between
various CR models.
With all four LEP experiments combined, it is likely that the
uncertainty from CR can be significantly reduced from its
current values.

The analysis of BEC effects is in a more advanced stage.
All four LEP collaborations are currently using an event
mixing method where data from two semi-leptonic \qqlns\
events are mixed (without the lepton) and compared to
data from genuine \qqqqs\ events.
In a rigorously model--independent test, these two samples
should look identical if there is no BEC present between the
decay products of different W bosons.
In order to limit the possible size of the effect allowed, 
however, a similar analysis must be performed on different 
Monte Carlo models of BEC.
Currently, all four LEP collaborations see no evidence for
inter-W BEC effects at the limits of their statistical power
to observe them.
By combining these results and evaluating a range of BEC
models, the systematic uncertainty from this source should
also be reduced for the final LEP results.

\begin{figure}
  \begin{center}
    \epsfig{file=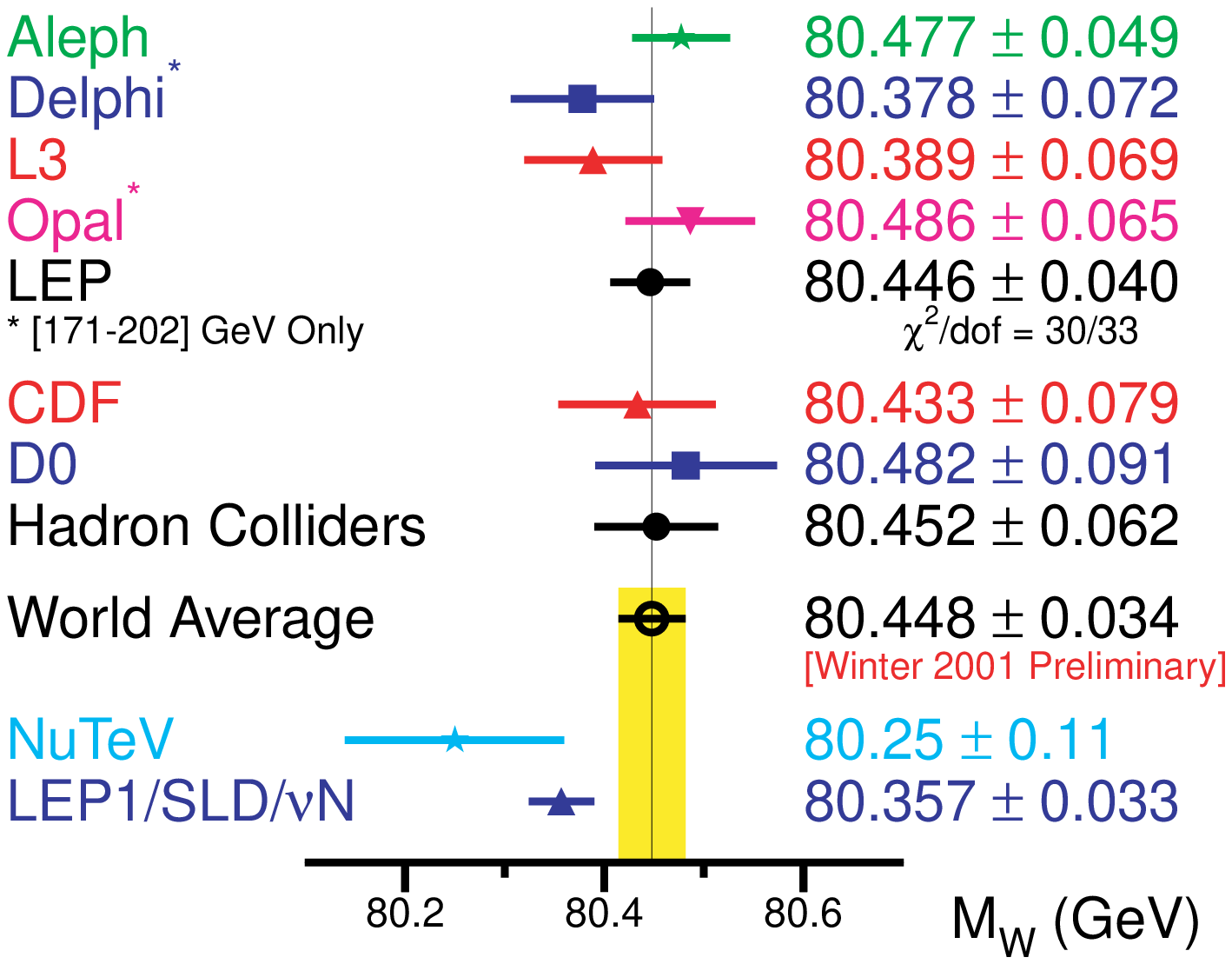,width=5.25cm}
    \hspace{0.5cm}
    \epsfig{file=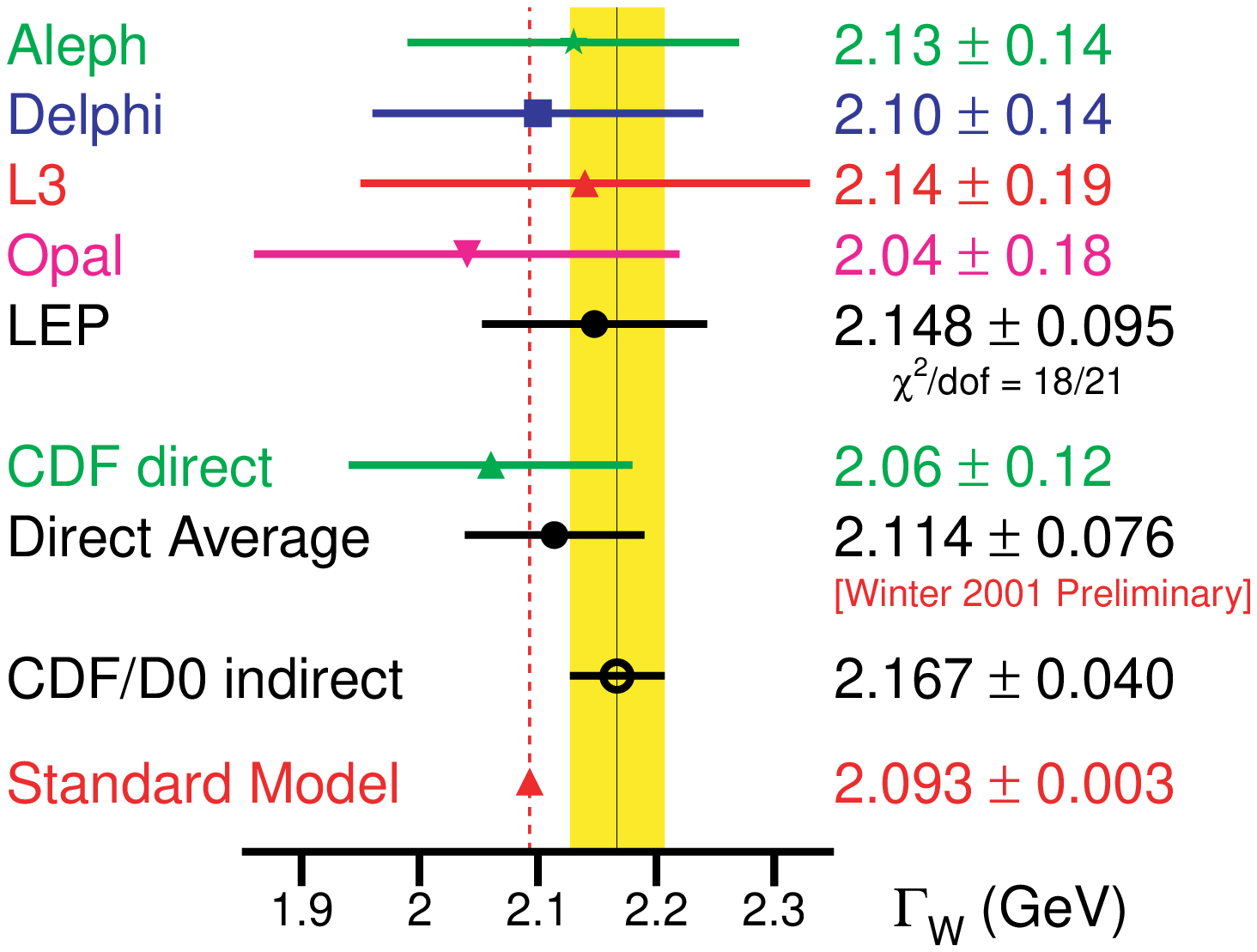,width=5.25cm}
    \caption{\it
      World results on the mass and width of the W boson.
      \label{fig:mwgw}
      }
 \end{center}
\end{figure}

The current world knowledge of the W mass from \LEPII\
and the Tevatron are shown in Figure~\ref{fig:mwgw},
along with the indirect value inferred from other 
precision electroweak measurements.
While the current world average is dominated by \LEPII,
with the data expected at the Tevatron during \RunIIa,
the combined error on direct W mass measurements should
approach 20 MeV by around 2006.
\subsection{W width measurements}
The direct reconstruction method employed at \LEPII, as well
as the transverse mass method employed at the Tevatron, are
sensitive to the W width as well as the W mass.
In the standard mass analyses, the width is typically fixed
to the Standard Model expectation for a given mass value,
but it can equally well be allowed to be a second free parameter
in the fits.
The direct measurements of the W boson width are shown in
Figure~\ref{fig:mwgw} along with the indirect width evaluation
mentioned earlier and the Standard Model expectation.
It should be noted that not all LEP data has been analyzed,
and D0 has yet to produce a result for the direct width
measurement.
With these improvements, the precision on the direct width
measurements should actually be competitive with the indirect
width from the Tevatron.
%
%
\section{Conclusions}
\begin{figure}
  \begin{center}
    \epsfig{file=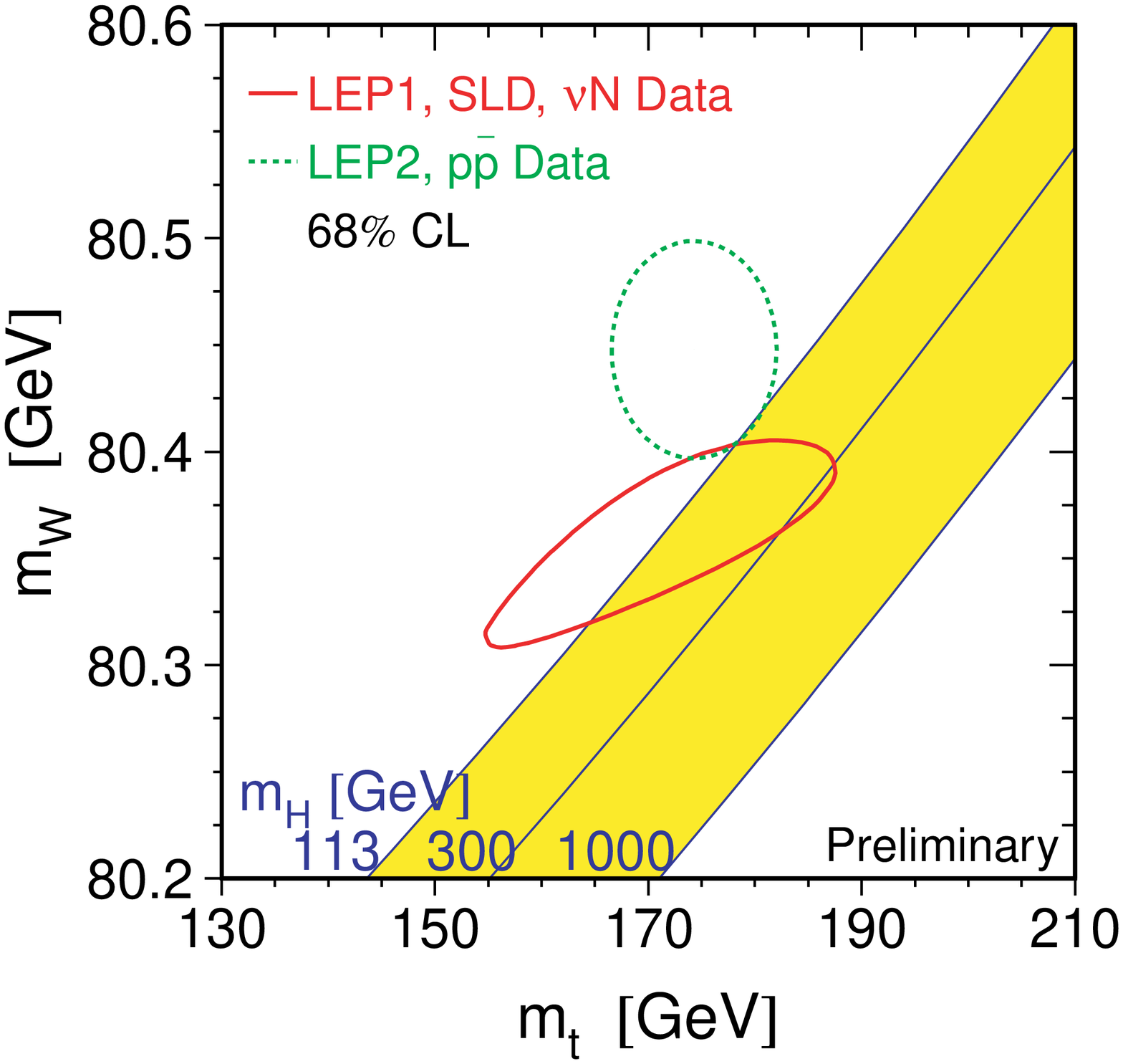,width=5.25cm}
    \hspace{0.5cm}
    \epsfig{file=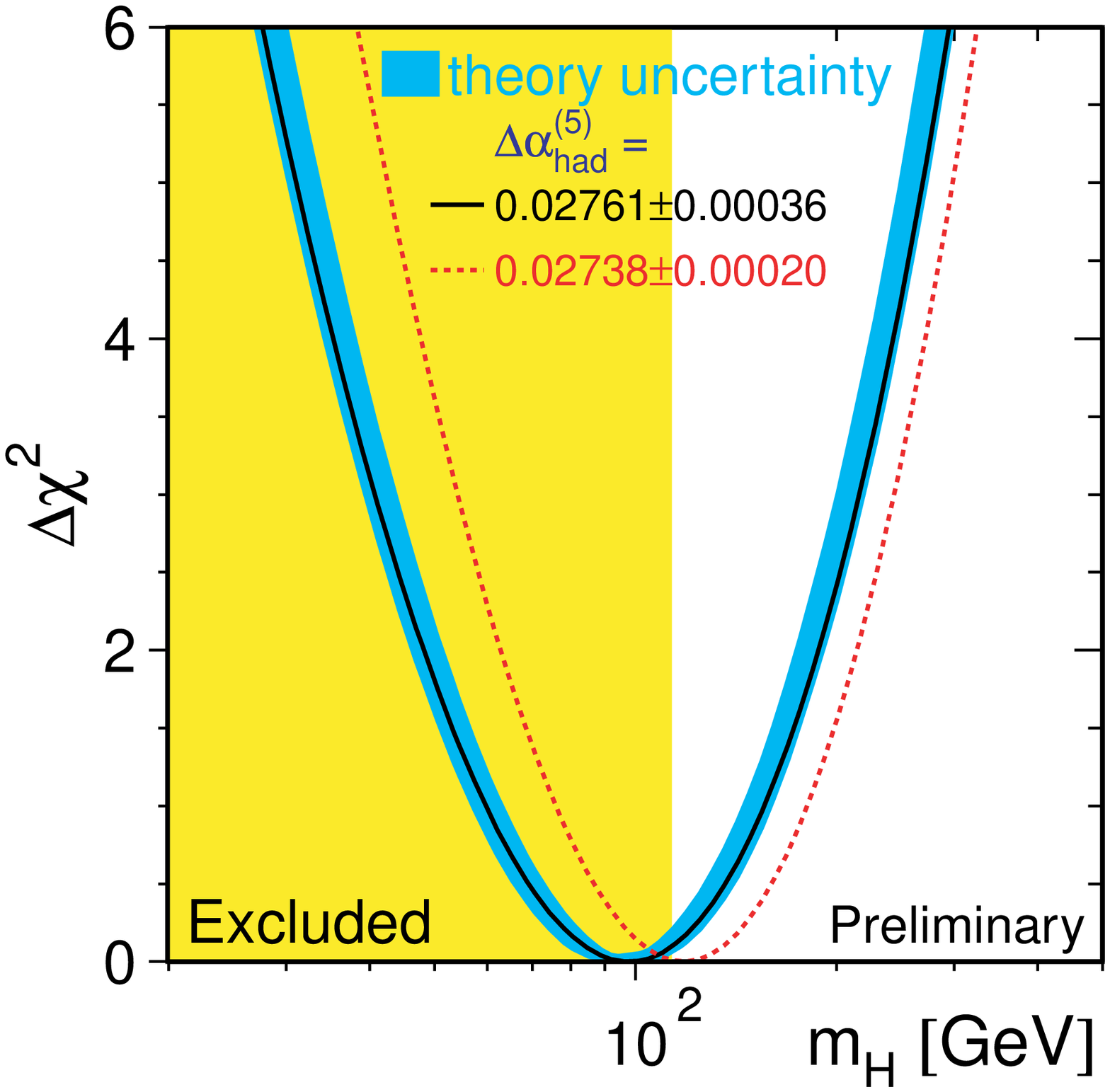,width=5.25cm}
    \caption{\it
      World combined electroweak results.
      \label{fig:worldew}
      }
 \end{center}
\end{figure}

The combined electroweak data is often summarized as shown
in Figure~\ref{fig:worldew}.
The first plot in this figure shows \Mw\ vs. \Mt\ both for
the direct measurements, the indirect electroweak data, and
the Standard Model prediction as a function of the Higgs mass.
It is clear that a light Higgs is preferred by the data, and
the direct measurements are starting to stray into a region
more favorable for Supersymmetric theories.
It can also be seen from this plot that significant improvements
in the uncertainty on \Mw\ will not be very useful if they
are not accompanied by comparable improvements in \Mt.

Another way of looking at this is in terms of the sensitivity
toward the remaining unknown parameter in the Standard Model,
the Higgs mass.
Several groups have produced simple parameterized versions
of the functional relationships predicted by the Standard
Model between derived quantities like \Mw\ or \sinw\
and the input parameters like \Mt\ and \Mh\cite{bib:poorzf}.
Using these tools, it can be shown that after \RunIIa, the
constraint on the Higgs mass arising from the measurement of
the W mass is improved in equal parts by reducing the uncertainty
on \Mw, but also by reducing the uncertainty on \Mt which
is necessary to calculate the Standard Model prediction.
It can also be seen that the constraint arising from the W mass
will match the precision of that coming from \sinw, and with
greatly reduced sensitivity to the running of the fine structure
constant\cite{bib:m01-ew}.

In summary, the past five years have seen some remarkable improvements
in the knowledge of the W boson properties from the LEP collaborations.
With around 650~\pbi\ recorded per experiment, world's best measurements
of the W boson branching fractions, gauge couplings, and mass have
been performed.
The next five years will see measurements of similar precision
performed at the Tevatron with the advent of \RunII.
The combined electroweak data continues to provide the most 
stringent constraint for any model of new physics, and improving
this knowledge is crucial step to the eventual goal of fully 
understanding the process of electroweak symmetry breaking.
%
%
\section{Acknowledgements}
The author would like to thank the organizers for the kind invitation 
to participate in this conference, as well as a warm personal thanks 
to the local organizing committee for their efforts to make this event 
pleasant for everybody.
The author would also like to thank the members of the
Aleph, Delphi, L3, Opal, CDF, and D0 collaborations for their
assistance in preparing this paper.
%
%

%
\end{document}